\documentclass[sigconf,nonacm]{acmart}

\AtBeginDocument{%
  \providecommand\BibTeX{{%
    \normalfont B\kern-0.5em{\scshape i\kern-0.25em b}\kern-0.8em\TeX}}}
    
 \usepackage{booktabs}
 \usepackage{multirow}
 \usepackage{graphicx}

\begin{document}

\title{Evaluating Information Retrieval Systems for Kids}
\author{Ashlee Milton}
\email{ashleemilton@u.boisestate.edu}
\affiliation{%
  \institution{PIReT -- People and Information Research Team \\Boise State University, Boise, Idaho, USA}
}

\author{Maria Soledad Pera}
\email{solepera@boisestate.edu}
\affiliation{%
  \institution{PIReT -- People and Information Research Team \\Boise State University, Boise, Idaho, USA}
}

\fancyhead{}


\begin{abstract}
Evaluation of information retrieval systems (IRS) is a prominent topic among information retrieval researchers--mainly directed at a general population. Children require unique IRS and by extension different ways to evaluate these systems, but as a large population that use IRS have largely been ignored on the evaluation front. In this position paper, we explore many perspectives that must be considered when evaluating IRS; we specially discuss problems faced by researchers who work with children IRS, including lack of evaluation frameworks, limitations of data, and lack of user judgment understanding.
\end{abstract}



\maketitle
\let\thefootnote\relax\footnotetext{"Copyright © 2020 for this paper by its authors. Use permitted under Creative Commons License Attribution 4.0 International (CC BY 4.0)."}

\section{The issue}
Evaluation in information retrieval (\textbf{IR}) remains a core research interest. Unlike most disciplines, evaluation of information retrieval systems (\textbf{IRS}), including popular ones like recommender systems and search engines, is not limited to effectiveness, as IR strives to give users what makes them happy not necessarily what they asked for \cite{callankeynote}. IR evaluation is an ongoing topic of interest among researchers and practitioners, with the focus being on metrics and their applicability for different tasks (e.g., multilingual IR), different domain (e.g., medical and legal), and applicability concepts like bias and fairness \cite{valcarce2018topn, anchez2018new, mehrotra2018fair, canamares2020offline, kelly2019clef}. The target users of these explorations on evaluation have been a traditional user \cite{callankeynote}. Thus, the frameworks and benchmarks used to evaluate IRS only account for a general user, who are not the only stakeholders.

It is our stance that the lack of evaluation frameworks for children's IRS is a problem that needs to be addressed by not only researchers working with children but the IR community. Evaluating IRS for children presents new challenges that have yet to be fully addressed. There have been a few evaluation \textit{frameworks} presented in recent years that attempt to create structure to enable assessment for kids' IRS \cite{landoni2019sonny, bilal2017towards}. Unfortunately, they are not always general enough to be applied to the varying IRS that are available for children. Even when these frameworks can be used, there is an overlying issue of \textit{data}. Data from children is hard to get in several respects including collecting, accessing, and storing due to laws protecting children \cite{federal1998children, rights1974privacy, rights2017privacy}. Even though some data may be accessible, how children \textit{judge} IRS differs from the general population \cite{hall2016five, jochmann2016interface, huibers2019relevance, wentzel2019evaluating}. We aim at showcasing (i) issues that ourselves and other researchers and practitioners face with data, user judgments, and frameworks, as well as (ii) the fact that evaluation that involves protected users is an issue with bigger implications that requires attention and must be informed by different areas of the IR community and beyond.

\section{The trouble with frameworks}
Existing frameworks have set out to make the best of the data that is available to evaluate children's IRS. The one presented in \cite{landoni2019sonny} brings context to evaluating and designing children's search systems with their four pillars: search strategy, user group, environment, and task. While originally designed for search systems, it is general enough to be used for context in other IR tasks \cite{downs2020guiding, downs2020kidspell}. However, the proposed framework does not explicitly address the limitations that arise due to lack of data. In \cite{bilal2017towards}, the authors present a framework to address the issue of relevance judgement with children regarding search. This framework is task-specific and would need manipulation to work with other IR tasks and areas. A useful framework that lends itself to IRS evaluation and that does not require data from children is the Cranfield paradigm \cite{cleverdon1967cranfield}. Cranfield uses known suitable resources as ground truth. However, it has been dropped and deemed outdated when evaluating IRS for general populations in favor of more state-of-the-art alternatives. The latter tend to require large amounts of data that is not always available with children making them not practical. Works like \cite{voorhees2019evolution}, have defended the Cranfield paradigm as a viable framework, but the IR community is divided on this issue. In the end, existing frameworks are a helpful foundation but each has constraints that render them insufficient for the IR community at large.

\section{The woes of data}
Due to privacy laws like the Children's Online Privacy Protection Act, Family Educational Rights and Privacy Act, and General Data Protection Regulation \cite{rights1974privacy, rights2017privacy, federal1998children}, children's data is highly protected. While these safeguards are of the upmost importance to keep children safe online, it makes collecting or finding children's data difficult. Thus, unless researchers have a means of gathering the needed data within the bounds of the child privacy laws themselves, it is impossible to get such data. Even then, there are additional rules as to what  can be collected (e.g., demographics), and how it must be stored. These stringent rules (i) make it extremely burdensome to share data and (ii) can lead to insufficient data for evaluating IRS for children.

\section{The ambiguity of judgment}
Existing ground truth may be misleading--it is not always what it seems. Studies exploring children's behavior as they interact with IRS and evaluation strategies for kid-friendly IRS~\cite{pera2019little, hall2016five} revealed that young users do not act in the same manners as adults do when interacting with or evaluating IRS \cite{bilal2002differences, hall2016five}. For example, kids tend to click on the first search result on a search result page regardless of its relevance \cite{TorresQueryChildren, gwizdka2016search}. Naturally, the thought is to use the child's clicked link as the relevant result but since they tend to favor the first result, regardless of relevance, does it work as ground truth? Even if you ask children what their judgments are, instead of trying to infer them, you can still have problems. Consider when asking kids to rate on a standard 1 to 5 scale, studies have shown they tend to only rate 4's or 5's regardless of what they think \cite{hall2016five}. Behaviors like these make it hard to define what the ground truth of collected data is and how applicable it is for conducting offline evaluations of IRS.

\section{The unknown}
The current state of evaluating IRS for children is in its infancy and it is indeed a complex undertaking driven by multiple perspectives \cite{murgia2019seven}. There is not general framework that can be used consistently and is accepted by the community as a whole; there is no reliable and/or standard way to obtain data for evaluation; and ground truth requires a unique perspective of relevance and that is just not the case when it comes to IRS for children. These issues showcase not only the importance of developing frameworks without the need for massive amounts of data but also why involving the larger community to create it is key. The reason for engaging with the IR community (and beyond) is two-fold. First, if the community is involved, they will become aware of the issues attached to the development and evaluation of IRS for children. Second, the researchers and practitioners can bring in their experiences on evaluation, especially from other areas working with protected populations. Working together we can learn from each other and hopefully come up with ways to facilitate the development of evaluation in different areas of study and bring the issues of evaluation of IRS for kids into the spotlight.

\bibliographystyle{ACM-Reference-Format}
\bibliography{References}


\begin{thebibliography}{00}


\ifx \showCODEN    \undefined \def \showCODEN     #1{\unskip}     \fi
\ifx \showDOI      \undefined \def \showDOI       #1{{\tt DOI:}\penalty0{#1}\ }
  \fi
\ifx \showISBNx    \undefined \def \showISBNx     #1{\unskip}     \fi
\ifx \showISBNxiii \undefined \def \showISBNxiii  #1{\unskip}     \fi
\ifx \showISSN     \undefined \def \showISSN      #1{\unskip}     \fi
\ifx \showLCCN     \undefined \def \showLCCN      #1{\unskip}     \fi
\ifx \shownote     \undefined \def \shownote      #1{#1}          \fi
\ifx \showarticletitle \undefined \def \showarticletitle #1{#1}   \fi
\ifx \showURL      \undefined \def \showURL       #1{#1}          \fi

\bibitem{bilal2017towards}
{Dania Bilal} {and} {Meredith Boehm}. 2017.
\newblock \showarticletitle{Towards new methodologies for assessing relevance
  of information retrieval from web search engines on children’s queries}.
\newblock {\em Qualitative and Quantitative Methods in Libraries\/} {2}, 1
  (2017), 93--100.
\newblock


\bibitem{bilal2002differences}
{Dania Bilal} {and} {Joe Kirby}. 2002.
\newblock \showarticletitle{Differences and similarities in information
  seeking: children and adults as Web users}.
\newblock {\em Information processing \& management\/} {38}, 5 (2002),
  649--670.
\newblock


\bibitem{callankeynote}
{Jamie Callan}. 2020.
\newblock Better Representation of Search Tasks.
\newblock Available at:
  \url{https://www.youtube.com/watch?v=eHJTkFUxJgg&feature=youtu.be&t=18047}.
  (2020).
\newblock
\newblock
\shownote{(accessed May 6).}


\bibitem{canamares2020offline}
{Roc{\'\i}o Ca{\~n}amares}, {Pablo Castells}, {and} {Alistair Moffat}. 2020.
\newblock \showarticletitle{Offline evaluation options for recommender
  systems}.
\newblock {\em Information Retrieval Journal\/} (2020), 1--24.
\newblock


\bibitem{cleverdon1967cranfield}
{Cyril Cleverdon}. 1967.
\newblock \showarticletitle{The Cranfield tests on index language devices}. In
  {\em Aslib proceedings}. MCB UP Ltd.
\newblock


\bibitem{federal1998children}
{Federal~Trade Commission} {and} {others}. 1998.
\newblock \showarticletitle{Children’s online privacy protection act of
  1998}.
\newblock  (1998).
\newblock


\bibitem{downs2020kidspell}
{Brody Downs}, {Oghenemaro Anuyah}, {Aprajita Shukla}, {Jerry Fails},
  {Maria~Soledad Pera}, {Katherine~Landau Wright}, {and} {Casey Kennington}.
  2020a.
\newblock \showarticletitle{KidSpell: A Child-Oriented, Rule-Based, Phonetic
  Spellchecker}. In {\em Proceedings of the The 11th International Conference
  on Language, Resources, and Evaluation} {\em (LREC ’20)}.
\newblock


\bibitem{downs2020guiding}
{Brody Downs}, {Aprajita Shukla}, {Mikey Krentz}, {Maria~Soledad Pera}, {Casey
  Kennington}, {Jerry Fails}, {and} {Katherine~Landau Wright}. 2020b.
\newblock \showarticletitle{Guiding the Selection of Child Spellchecker
  Suggestions using Audio and Visual Cues}. In {\em Proceedings of the The 19th
  International Conference on Interaction Design and Children} {\em (IDC
  ’20)}. Association for Computing Machinery, New York, NY, USA.
\newblock


\bibitem{TorresQueryChildren}
{Sergio Duarte~Torres}, {Djoerd Hiemstra}, {and} {Pavel Serdyukov}. 2010.
\newblock \showarticletitle{Query Log Analysis in the Context of Information
  Retrieval for Children}. In {\em Special Interest Group on Information
  Retrieval}. ACM, 847--848.
\newblock


\bibitem{gwizdka2016search}
{Jacek Gwizdka}, {Preben Hansen}, {Claudia Hauff}, {Jiyin He}, {and} {Noriko
  Kando}. 2016.
\newblock \showarticletitle{Search as learning (SAL) workshop 2016}. In {\em
  39th International SIGIR conference on Research and Development in
  Information Retrieval}. ACM, 1249--1250.
\newblock


\bibitem{hall2016five}
{Lynne Hall}, {Colette Hume}, {and} {Sarah Tazzyman}. 2016.
\newblock \showarticletitle{Five Degrees of Happiness: Effective Smiley Face
  Likert Scales for Evaluating with Children}. In {\em Proceedings of the The
  15th International Conference on Interaction Design and Children} {\em (IDC
  ’16)}. Association for Computing Machinery, New York, NY, USA, 311–321.
\newblock
\showISBNx{9781450343138}
\showDOI{%
\url{http://dx.doi.org/10.1145/2930674.2930719}}


\bibitem{huibers2019relevance}
{Theo Huibers} {and} {Thijs Westerveld}. 2019.
\newblock \showarticletitle{Relevance and utility in an educational search
  environment}. In {\em 3rd International and Interdisciplinary Perspectives on
  Children \& Recommender and Information Retrieval Systems, KidRec 2019: what
  does good look like?}
\newblock


\bibitem{jochmann2016interface}
{Hanna Jochmann-Mannak}, {Leo Lentz}, {Theo Huibers}, {and} {Ted Sanders}.
  2016.
\newblock \showarticletitle{How Interface Design and Search Strategy Influence
  Children's Search Performance and Evaluation}.
\newblock In {\em Web Design and Development: Concepts, Methodologies, Tools,
  and Applications}. IGI Global, 1332--1379.
\newblock


\bibitem{kelly2019clef}
{Liadh Kelly}, {Lorraine Goeuriot}, {Hanna Suominen}, {Mariana Neves},
  {Evangelos Kanoulas}, {Rene Spijker}, {Leif Azzopardi}, {Dan Li}, {Jo{\~a}o
  Palotti}, {Guido Zuccon}, {and} {others}. 2019.
\newblock \showarticletitle{CLEF ehealth 2019 evaluation lab}. In {\em European
  Conference on Information Retrieval}. Springer, 267--274.
\newblock


\bibitem{landoni2019sonny}
{Monica Landoni}, {Davide Matteri}, {Emiliana Murgia}, {Theo Huibers}, {and}
  {Maria~Soledad Pera}. 2019.
\newblock \showarticletitle{Sonny, Cerca! evaluating the impact of using a
  vocal assistant to search at school}. In {\em International Conference of the
  Cross-Language Evaluation Forum for European Languages}. Springer, 101--113.
\newblock


\bibitem{mehrotra2018fair}
{Rishabh Mehrotra}, {James McInerney}, {Hugues Bouchard}, {Mounia Lalmas},
  {and} {Fernando Diaz}. 2018.
\newblock \showarticletitle{Towards a Fair Marketplace: Counterfactual
  Evaluation of the Trade-off between Relevance, Fairness \& Satisfaction in
  Recommendation Systems}. In {\em Proceedings of the 27th ACM International
  Conference on Information and Knowledge Management} {\em (CIKM ’18)}.
  Association for Computing Machinery, New York, NY, USA, 2243–2251.
\newblock
\showISBNx{9781450360142}
\showDOI{%
\url{http://dx.doi.org/10.1145/3269206.3272027}}


\bibitem{murgia2019seven}
{Emiliana Murgia}, {Monica Landoni}, {Theo Huibers}, {Jerry~Alan Fails}, {and}
  {Maria~Soledad Pera}. 2019.
\newblock \showarticletitle{The Seven Layers of Complexity of Recommender
  Systems for Children in Educational Contexts}. In {\em Workshop on
  Recommendation in Complex Scenarios, co-located with ACM RecSys}. 5--9.
\newblock
\showDOI{%
\url{http://dx.doi.org/Vol-2449/paper1.pdf}}


\bibitem{pera2019little}
{Maria~Soledad Pera}, {Emiliana Murgia}, {Monica Landoni}, {and} {Theo
  Huibers}. 2019.
\newblock \showarticletitle{With a Little Help from My Friends: Use of
  Recommendations at School}.
\newblock  (2019).
\newblock


\bibitem{rights1974privacy}
{Family~Educational Rights}. 1974.
\newblock Privacy Act or 1974.
\newblock   (1974).
\newblock


\bibitem{rights2017privacy}
{Family~Educational Rights}. 2017.
\newblock Privacy Act of 2017.
\newblock   (2017).
\newblock


\bibitem{anchez2018new}
{Pablo S\'{a}nchez}, {Rus~M. Mesas}, {and} {Alejandro Bellog\'{\i}n}. 2018.
\newblock \showarticletitle{New Approaches for Evaluation: Correctness and
  Freshness: Extended Abstract}. In {\em Proceedings of the 5th Spanish
  Conference on Information Retrieval} {\em (CERI ’18)}. Association for
  Computing Machinery, New York, NY, USA, Article 14, 2 pages.
\newblock
\showISBNx{9781450365437}
\showDOI{%
\url{http://dx.doi.org/10.1145/3230599.3230614}}


\bibitem{valcarce2018topn}
{Daniel Valcarce}, {Alejandro Bellog\'{\i}n}, {Javier Parapar}, {and} {Pablo
  Castells}. 2018.
\newblock \showarticletitle{On the Robustness and Discriminative Power of
  Information Retrieval Metrics for Top-N Recommendation}. In {\em Proceedings
  of the 12th ACM Conference on Recommender Systems} {\em (RecSys ’18)}.
  Association for Computing Machinery, New York, NY, USA, 260–268.
\newblock
\showISBNx{9781450359016}
\showDOI{%
\url{http://dx.doi.org/10.1145/3240323.3240347}}


\bibitem{voorhees2019evolution}
{Ellen~M Voorhees}. 2019.
\newblock \showarticletitle{The evolution of cranfield}.
\newblock In {\em Information Retrieval Evaluation in a Changing World}.
  Springer, 45--69.
\newblock


\bibitem{wentzel2019evaluating}
{SD Wentzel}. 2019.
\newblock {\em Evaluating Information Retrieval Systems for Children in an
  Educational Context}.
\newblock {B.S.} thesis. University of Twente.
\newblock


\end{thebibliography}

\end{document}